# Single pulse all-optical toggle switching of magnetization without Gd: The example of $Mn_2Ru_xGa$

C. Banerjee, N. Teichert, K. Siewierska, Z. Gercsi, G. Atcheson, P. Stamenov, K. Rode, J. M. D. Coey and J. Besbas*

[1] CRANN, AMBER and School of Physics, Trinity College, Dublin 2, Ireland

**Energy-efficient control of magnetization without the help of a magnetic field is a key goal of spintronics[1,2]. Purely heat-induced single-pulse all-optical toggle switching has been demonstrated, but so far only in Gd based amorphous ferrimagnet films[3–6]. In this work, we demonstrate toggle switching in the half-metallic compensated ferrimagnetic Heusler alloys $Mn_2Ru_xGa$, which have two crystallographically-inequivalent Mn sublattices[7]. Moreover, we observe the switching at room temperature in samples that are immune to external magnetic fields in excess of 1 T, provided they exhibit compensation above room temperature. Observations of the effect in compensated ferrimagnets without Gd challenges our understanding of all-optical switching. The dynamic behavior indicates that $Mn_2Ru_xGa$ switches in 2 ps or less. Our findings widen the basis for fast optical switching of magnetization and break new ground for engineered materials that can be used for nonvolatile ultrafast switches using ultrashort pulses of light.**

*besbasj@tcd.ie

Driven by the demands for high speed, low cost and high-density magnetic recording, research in spintronics has always sought insight into new classes of magnetic materials and devices that show efficient and reproducible magnetization switching. In this respect, interest in the magnetic properties of antiferromagnetically coupled sub-lattice systems has gained momentum in the last decade. The total or partial cancellation of the moments makes these systems insensitive to stray magnetic fields, and the interaction between the sublattice moments introduces phenomena that are absent in conventional ferromagnets, opening new opportunities for magnetic recording and information processing[2,8–10].

An efficient way of controlling magnetism is to use ultrashort laser pulses[1,11]. XMCD investigations in 2011 by Radu *et al.*[3] of the dynamics of the Gd and Fe atomic moments in a thin layer of amorphous ferrimagnetic $Gd_{25}Fe_{65.6}Co_{9.4}$ after a 50 fs laser pulse, revealed a transient parallel alignment of the moments that was the precursor of switching. They were followed by the discovery of single-pulse all-optical toggle switching of the magnetization in the same material by Ostler *et al.*[12]. A general basis for fast all-optical switching in multi-sublattice magnets was then proposed by Mentink *et al.*[13]. Amorphous $Gd_x(Fe,Co)_{100-x}$ with $x \approx 25$ is a metallic ferrimagnet with localized 4*f*-shell magnetic moments on the Gd sublattice and delocalized 3*d*-band moments on the Fe-Co sublattice. Upon excitation by a femtosecond laser pulse, the Fe-Co undergoes sub-picosecond demagnetization leading to practically complete loss of the ordered 3*d* shell magnetization, an effect that had been originally observed in ferromagnetic nickel[14]. Concomitantly, the Gd atoms experience a slower loss of magnetic alignment, with partial transfer of angular momentum from the Gd *f* shell to the FeCo *d* shell[3], entailing a transient parallel alignment of the moments of the demagnetizing Gd and the re-magnetizing FeCo that ultimately leads to magnetization toggle switching on a picosecond timescale[3,12]. As the suggested mechanism for single pulse all optical switching (SP-AOS) relies on an ultrafast interplay between two inequivalent spin sublattices, one with a slower response to the laser (the Gd 4*f* electrons) and the other with a faster one (the Fe 3*d* electrons), subsequent researches on SP-AOS concentrated on rare-earth based ferrimagnets[15]. In these switching measurements, it is useful to distinguish the very short timescale on which the future direction of the net magnetization is decided, and the longer timescale necessary for the equilibrium magnetization to be established and respond to an external magnetic field. In practice helicity-independent SP-AOS has only been demonstrated in ferrimagnetic $Gd_x(Fe,Co)_{100-x}$ thin films[3], $Gd_x(Fe,Co)_{100-x}$ spin valves[5] and in synthetic Gd/Co ferrimagnets[4]; It has not been seen in other rare-earth based ferrimagnets such as amorphous $Tb_{27}Co_{73}$[16,17] where the 4*f* electrons experience strong spin-orbit coupling. Its thermal origin is established by the independence of the effect on the polarization and helicity of the light[3,12], and the equivalent effect produced by pulses of hot electrons[18]. A related phenomenon has been reported in ferrimagnetic TbFeCo, however, under specific structural conditions[19] and in ferromagnetic Pt/Co/Pt structures, when the laser spot size matches that of the ferromagnetic domains[20]. A different type of single-pulse, non-thermal, non-toggle switching has been reported with linearly-polarized light in insulating Co-doped yttrium iron garnet[21].

In this work, we present a new, rare-earth-free, ferrimagnet that exhibits SP-AOS where the two sublattices should not according to the prevalent thermodynamical models[13] have drastically different response time to laser excitation. We report all-optical toggle switching in the ferrimagnetic Heusler alloys $Mn_2Ru_xGa$ (MRG)[7] where both magnetic sublattices are composed of manganese, and establish MRG as a versatile alternative to $Gd_x(FeCo)_{1-x}$ for SP-AOS applications. In MRG, the Mn atoms occupy two inequivalent sub-lattices at Wyckoff positions 4*a* and 4*c* in the cubic $F\acute{4}3m$ structure (See supplementary Fig. S1a), with antiferromagnetic intersublattice coupling[7]. At low temperature the magnetization of the Mn(4*c*) sublattice is dominant, but as temperature increases the magnetization of the Mn(4*c*) falls faster than that of the Mn(4*a*) sublattice, leading to a compensation temperature $T_{comp}$ where the two are equal and opposite as the coercivity tends to diverge when the net magnetization crosses zero[22]. The value of $T_{comp}$ can be varied by changing the Ru concentration *x*, so it is possible to make MRG peculiarly insensitive to external magnetic fields by decreasing its magnetisation[23]. The electronic structures of the two sublattices are different. MRG has a spin gap ~1 eV close to the Fermi energy, which led to its identification as the first example of a half-metallic ferrimagnet[7]; the Mn(4*c*) electrons have a high spin-polarized density of states whereas that of the Mn(4*a*) electrons is much lower. The unusual electronic structure accounts for an anomalous Hall effect (AHE) that is greater than those seen in common ferromagnets[23] and a strong magneto-optical Kerr effect (MOKE), even when the net magnetization vanishes at $T_{comp}$[23,24], because both AHE and MOKE probe mainly the spin-polarized conduction band associated with Mn in the 4c position. Domains can be directly imaged in the Kerr microscope, regardless of the net magnetization [25].

In our experiments, we investigated SP-AOS in 19 MRG thin films having different Ru contents with $T_{comp}$ above or below room temperature (RT). The films are deposited on MgO (100) substrates, which leads to a slight tetragonal distortion of the cubic XA-type structure (from space group 216, $F\acute{4}3m$ to space group 119, $I\acute{4}m2$), which is responsible for the perpendicular magnetic anisotropy of the MRG films. Optical pulses of 800 nm wavelength and about 200 fs duration were generated by a mode-locked Ti-sapphire laser seeding a 1 kHz amplifier. Figure 1 displays the results of irradiating a $Mn_2Ru_{1.0}Ga$ film by a single 200 fs pulse with a Gaussian intensity profile, as observed by *ex situ* Kerr microscopy. Here the light or dark contrast indicates an orientation of the Mn(4*c*) sublattice into or out of the plane. For either initial magnetization direction, a single laser pulse of sufficient intensity will switch the magnetization direction in the irradiated area (The elliptical shape of the switched domain is caused by astigmatism of the focusing lens). Pulses where the average energy density is sub-threshold leave the magnetization unchanged, except just at the centre, where the intensity exceeds threshold (Fig 1a). The whole irradiated spot is switched at 1.5 µJ (Fig 1b), but at 3 µJ a multidomain pattern appears in the center of the irradiated zone (Fig 1c), where the temperature of the film has transiently exceeded the Curie temperature of the sample (~500 K)[7] leading to re-magnetization in sub-micron domains close in size to the resolution of the Kerr microscope. They are much smaller than the ~ 100 µm domains normally observed at room temperature during the reversal process after saturating the magnetization[25]. It is established that such temperatures can be reached in equilibrium between the lattice and spin system in the very first picoseconds following optical excitation in transition metal compounds[14], after which the system re-magnetizes randomly in the stray field during the cool down. The multidomain pattern is directly surrounded by a ring-shaped switched domain, which shows that SP-AOS involves a significant transient demagnetization. The variation of the size of the switched domain area with increasing pulse energy has been employed to calculate the threshold fluence for switching (See Supplementary Information VI). Interestingly, we never observed SP-AOS in any MRG film having $T_{comp}$ below RT (see Supplementary Information V). We verified that the observed sequence of switching originates solely from laser induced heating, by repeating the experiment with circularly polarized laser pulses of opposite helicities as well as different directions of linear polarization with respect to the MRG crystallographic directions (not shown). The SP-AOS occurred identically in all cases, which eliminates the possibility of any contribution from magnetic circular dichroism[26] or from transient spin-orbit torques generated by the electric field. On further increasing the laser power to 5 μJ, the center of the irradiated spot on the film is ablated.

Figure 2 depicts the results of the irradiation with 1 to 5 successive laser pulses on $Mn_2Ru_{1.0}Ga$. The panels show different regions that were subjected to the given numbers of shots. Consistently, the irradiation by a series of laser pulses leads to a toggling of the direction of the magnetization, which was investigated for up to 12 consecutive pulses.

MRG possesses a low net magnetization and a high anisotropy field. Therefore, the coercive field of the films usually exceeds 0.2 T and can reach values as high as 10 T[23] if the temperature is very close to $T_{comp}$. It is interesting to see whether a highly-coercive sample can be switched by light at RT. Figure 3 shows the toggling of magnetization following a sequence of pulses in a film of $Mn_2Ru_{0.75}Ga$ with coercivity exceeding 1 T. That sample could not be saturated in our electromagnet and it was therefore measured in its virgin state, which is characterized by a distribution of magnetic domains with a predominance of magnetization directed toward the substrate. Toggling of each of the individual domains by the light pulse is observed even though the sample is insensitive to an external magnetic field of 1 T. The threshold fluence for this sample was approximately one third of that of $Mn_2Ru_{1.0}Ga$. The SP-AOS observed in samples with compensation temperature close to RT is particularly important for three reasons: 1) the threshold fluence required for switching is small, potentially enabling energy-efficient applications in the future[27]; 2) the coercivity of MRG diverges close to $T_{comp}$, which makes the magnetic state impervious to external magnetic fields; 3) switching of micron-sized domains is possible, which are much smaller than the laser spot size.

Next we turn to the dynamics of the excitation and reversal processes. The magnetism in MRG originates from the 3*d* moments of Mn(4*a*) and Mn(4*c*) sublattices which are antiferromagnetically coupled. The femtosecond laser pulse is expected to disrupt the inter-site exchange (< 0.1 eV) and rapidly destroy the magnetic order; while the intra-atomic, on-site exchange that depends on stronger Coulomb interactions (3 – 5 eV) should not be completely destroyed. The aftermath of the pulse therefore involves re-establishment of magnetic order from the atomic moments, which in a ferrimagnet could include effects of angular momentum transfer between sites. To investigate this possibility, we have studied the magnetization dynamics using time-resolved polar MOKE (TR-MOKE) in the two-color collinear pump probe geometry. In this part of the study, we compare two samples, $Mn_2Ru_{1.0}Ga$ and $Mn_2Ru_{0.65}Ga$. They have $T_{comp}$ of 390 K and 165 K respectively and their coercive fields at room temperature are similar (~ 460 mT), as shown in the optically measured hysteresis loops in Fig. 4a. The loops have opposite signs, as expected, because the Mn(4*c*) sublattice, which gives the dominant contribution to the MOKE signal, aligns parallel to the applied field below $T_{comp}$ and antiparallel above. Intense laser pulses of wavelength 800 nm were used as the pump beam to excite the magnetization dynamics, and the Mn(4c) sublattice magnetization was subsequently probed in a stroboscopic manner using weaker 400 nm pulses. A field of 500 mT was applied perpendicular to the films to ensure an identical initial state before each pump pulse. Figure 4b shows the TR-MOKE signal for different pump fluences for $Mn_2Ru_{0.65}Ga$, which does not switch because $T_{comp}$ is below RT. Following the laser excitation, the transient MOKE signal shows a step-like change, caused by the ultrafast destruction of the magnetic order as the electron temperature shoots up. Subsequently, the magnetization regains its initial state in tens or hundreds of ps, depending on laser fluence. This behavior is typical of ferromagnetic metals. It should be noted here that even though the MOKE response indicates full demagnetization of the Mn(4*c*) electrons, this does not mean we have transiently exceeded the Curie temperature in the lattice. At a fluence of 15.1 mJ $cm^{-2}$, a trace of magnetic order might still persist in the magneto-optically silent Mn(4*a*) sublattice, to ensure that when the system re-magnetizes, it does so uniformly. At greater fluences we observe a multi-domain state after laser irradiation, indicating that the system has been thermally demagnetized when the electrons re-establish thermal equilibrium with a lattice that is now above $T_c$.

On changing to $Mn_2Ru_{1.0}Ga$, we find a strong dependence of the TR-MOKE signal on laser fluence, which is quite different below and above the threshold fluence $F_{th}$ for SP-AOS (See Fig. $4c_1$ and $4c_2$ respectively). Below threshold, the behavior is like that of $Mn_2Ru_{0.65}Ga$ at similar fluence; the recovery takes about 10 ps, and an increase in the fast demagnetization signal is observed with increasing fluence (Fig 4*d*). Upon crossing the fluence threshold indicated by the yellow bar, the following new features appear in the signal: i) an increase in the pump fluence now leads to a *decrease* in the demagnetization amplitude at zero delay (Fig. 4c$_3$), contrary to the previous case and to $Gd_x(FeCo)_{1-x}$[28]. The behavior of the demagnetization amplitude with increasing pump fluence is seen in Fig. 4d, where it falls away from full demagnetization above $F_{th}$; ii) As the system relaxes, the signal undergoes a rapid partial recovery within 2 ps (See Fig. 4c$_3$), after which it reverses, at the point marked by the arrow. This anomaly, which has not been reported in $Gd_x(FeCo)_{1-x}$, appears in the sample compensating above room temperature and only at fluences where toggle switching is observed. There, at 2 ps, the film must have switched, in the short-time sense mentioned in the introduction, because at longer times the Mn(4*c*) sublattice continues to re-magnetize in a direction opposite to the one it had originally, before being turned back after 50 ps by the weak applied field and gradually recovering over the course of several hundred picoseconds. Extrapolating the initial slope to negative saturation gives a switching time, in the longer sense, of about 200 ps. The slope from 2 – 50 ps is comparable, but opposite in sign to that of the first sample (Fig 2b) and the timescale for recovery at high fluence is similar.

Another sample with $T_{comp}$ = 250 K, excited at 210 K or 230 K was found to behave similarly[29], and the time for thermalization of the electrons and the lattice is deduced there from a 4-temperature model to be 2 ps. Furthermore, Bonfiglio *et al.*[29] have shown that magnetic order is already beginning to be re-established in MRG within 1 ps, permitting efficient exchange scattering and transfer of angular momentum from one sublattice to the other, even at extremely short timescales. We speculate that for MRG most of the demagnetization is due to exchange of angular momentum between the sublattices, but a quite different wavelength of the probe pulse would be needed to reveal the behavior of the 4*a* sublattice and see whether the transient parallel alignment of the two subattice moments that is seen in XMCD in $Gd_x(FeCo)_{1-x}$[3] is also present in MRG. The question of where the local memory of the magnetization is stored just after the electrons are demagnetized by the laser pulse is germane to the explanation of the behavior shown in Fig 1 both below and above the switching threshold $F_{th}$. If not in a substantially slower demagnetization rate for the silent 4*a* sublattice, then the local angular momentum might be transiently parked in optical phonon modes, or the nuclear spins.

In summary, we have demonstrated single-pulse all-optical thermal switching in less than 2 ps in films of the half-metallic compensated Heusler ferrimagnet $Mn_2Ru_xGa$, where both magnetic sublattices are composed of manganese atoms, occupying different crystallographic sites. These results extend the scope of the phenomenon beyond a limited range of amorphous $Gd_x(Fe,Co)_{100-x}$ alloys with $x \approx 25$, where the magnetic sublattices are defined chemically. A comparison of the two systems is provided in Supplemental Information VIII. The Heusler alloys are a huge family, with an established body of knowledge about their magnetic and electronic properties that will allow us to advance our understanding of the SP-AOS phenomenon and design materials that can be the basis of future nonvolatile opto-magnetic switches. Beyond the newly-demonstrated quality of MRG as an opto-magnetic material, its large intrinsic spin-orbit torque, which relies on the absence of inversion symmetry of the Mn(4*c*) sublattice opens prospects for new multifunctionality[30,31]. Therefore, MRG and its chemically tailored successors offer the prospects of both new insights into condensed matter on a femtosecond timescale and new technological prospects taking advantage of ultra-fast control of the magnetic state without any reliance on a magnetic field.

**Methods:** MRG films with different Ru content were grown on MgO(001) substrates at 350°C by DC magnetron sputtering in a Shamrock system with a base pressure of $2\times10^{-8}$ Torr. They were co-sputtered from Ru and stoichiometric $Mn_2Ga$ targets. The Ru concentration was controlled by varying the $Mn_2Ga$ target plasma power while fixing the Ru power. The samples were then capped with a protective layer of 2 nm of $Al_2O_3$.

Femtosecond laser pulses were generated by Ti-sapphire laser seeding a 1 kHz amplifier with a Q-switched cavity. Their central wavelength was 800 nm and the pulse duration was about 200 fs. The amplifier can be operated in continuous mode where a train of pulses is generated at a repetition rate of 1 kHz or in single pulse mode where the emission of one single pulse can be externally triggered. In some cases, 400 nm laser pulses were obtained by second harmonic generation in a $\beta$-$BaB_2O_4$ crystal.

Prior to laser irradiation, the films were saturated at room temperature in the 1 T perpendicular magnetic field of our Evico Kerr microscope. Different areas of the films were then irradiated with several linearly polarized laser pulses of different powers, followed by ex-situ imaging of the final results in the Kerr microscope. For the imaging, a polarized beam is focused onto the sample using a microscope objective. A rotation in the polarization due to Kerr effect occurs in the reflected beam, which then passes through an analyzer, before reaching the camera. In order to increase the contrast, we keep the axis of the analyzer few degrees away from cross position in both directions and acquire two images. Subsequently, the images are subtracted to extract the final image.

For the dynamic measurements, the laser beam with wavelength 800 nm was split into a pump beam and a frequency-doubled probe beam at 400 nm. The intensity of the probe was kept low. Both pump and probe were linearly polarized and collinear. The spot sizes were measured to be about 150 and 70 μm respectively. The dynamical magneto-optic Kerr rotation was measured using a balanced detection scheme and acquired using a lock-in amplifier and a mechanical chopper at 500 Hz in the pump beam. The pump/probe delay was varied using a mechanical delay line.

**Figure 1| Single-pulse all-optical switching (SP-AOS) in $Mn_2Ru_{1.0}Ga$.** A film uniformly-magnetized out of the plane (top) or into the plane (bottom) is irradiated by a single 800 nm pulse focused onto a spot 150µm x 100µm. The pulse energy in **a** is 0.9 µJ, which only exceeds the switching threshold in a small region at the centre where the fluence is highest. The 1.5µJ pulse in **b** switches the whole irradiated area, whereas the 3µJ pulse in **c** heats the film at the centre of the spot above the Curie temperature, producing a fine multidomain pattern. The most intense pulses (5 µJ) lead to ablation of the film.

**Figure 2| Toggling of the magnetization in $Mn_2Ru_{1.0}Ga$.** Magnetization pattern as a function of the number of applied pulses. Pulse energy was 1.2 µJ.

**Figure 3| Toggling of magnetization in a high-coercivity $Mn_2Ru_{0.75}Ga$ film.** Repeated toggling of the micron-scale domain pattern of a virgin-state sample is observed with repeated pulses. There was a net imbalance of domains pointing in and out of the plane.

**Figure 4| Time Resolved Magnetization Dynamics in $Mn_2Ru_xGa$. a,** Hysteresis loops measured by MOKE in $Mn_2Ru_{0.65}Ga$ and $Mn_2Ru_{1.0}Ga$, which have compensation temperature below and above RT respectively. **b,** Transient Kerr signals of $Mn_2Ru_{0.65}Ga$ for different pump fluences. The variation of the Kerr signal is normalized to the total Kerr rotation at RT. **$c_1$,** Transient Kerr signal of $Mn_2Ru_{1.0}Ga$ for fluences below the switching threshold; **$c_2$**, similar data including fluences above threshold; **$c_3$**, A zoom of **$c_2$** in a shorter time window; the anomaly marked by the arrow is discussed in the text. **d,** Variation of the demagnetization amplitude at zero delay for different pump fluences. The yellow shaded region indicates the threshold fluence for switching. The solid line is a guide to eye.

**Author contributions:**

C. B., J. B. and J. M. D. C. designed the project. Experimental work was done by C. B., N. T. and J. B. Growth and characterization of the samples were carried out by G. A. and K. S. Numerical simulations were performed by Z. G. and J. B.; C. B., J. B., P. S., J. M. D. C and K. R. interpreted the data. All authors discussed the results. C. B., J. B., K. R. and J. M. D. C. wrote the paper.

**Acknowledgements:**

This project has received funding from Science Foundation Ireland through contracts 16/IA/4534 ZEMS and 12/RC/2278 AMBER and from the European Union's FET-Open research programme under grant agreement No 737038. C.B. is grateful to Irish Research Council for her post-doctoral fellowship. N. T. would like to acknowledge funding from the European Union's Horizon 2020 research and innovation programme under the Marie Skłodowska-Curie EDGE grant agreement No 713567.

The authors declare no competing financial interest.

**Figure 1|**

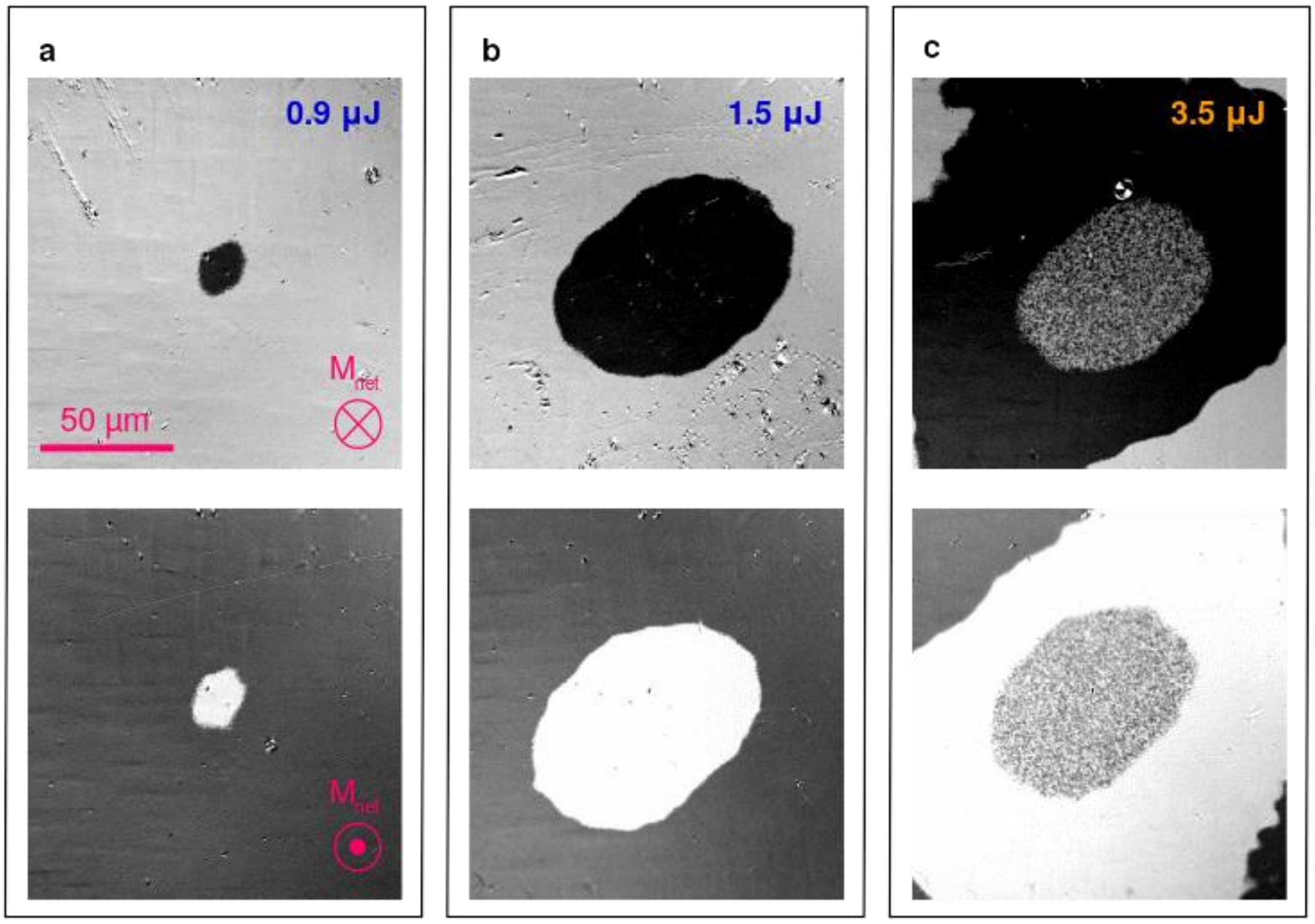

**Figure 2|**

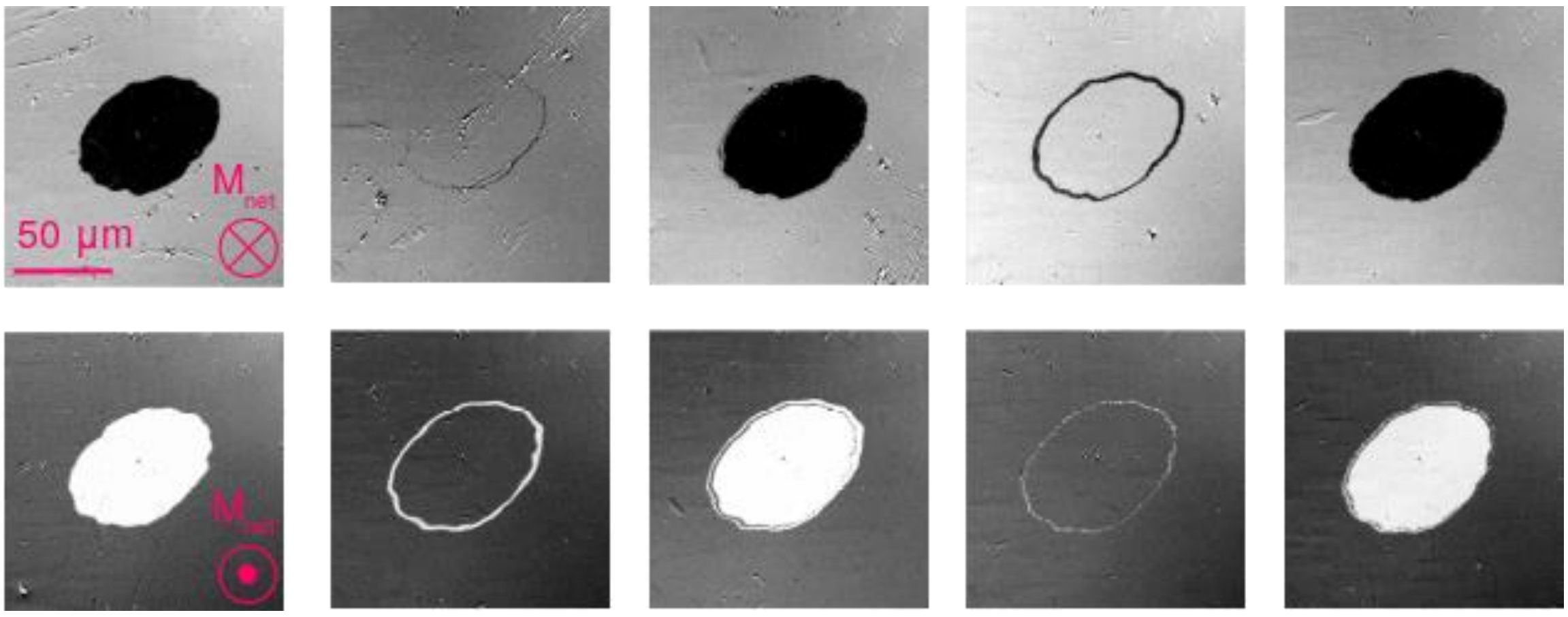

**Figure 3|**

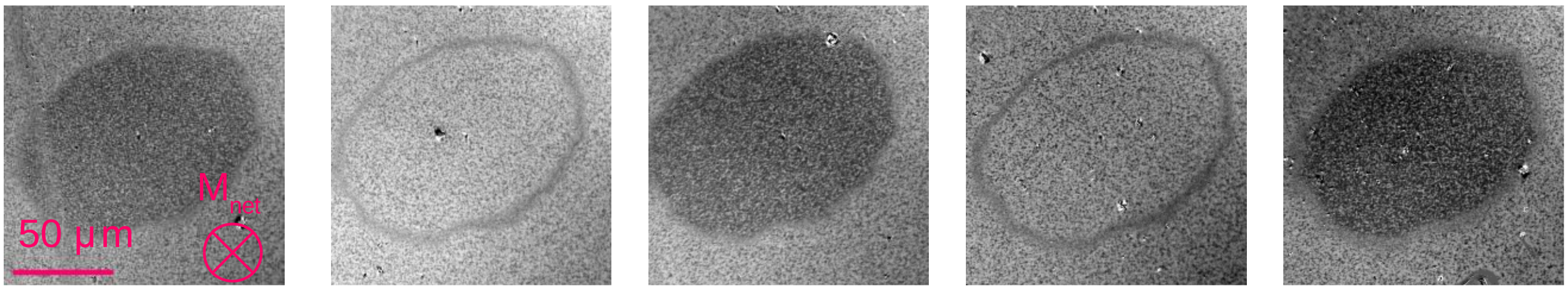

**Figure 4|**

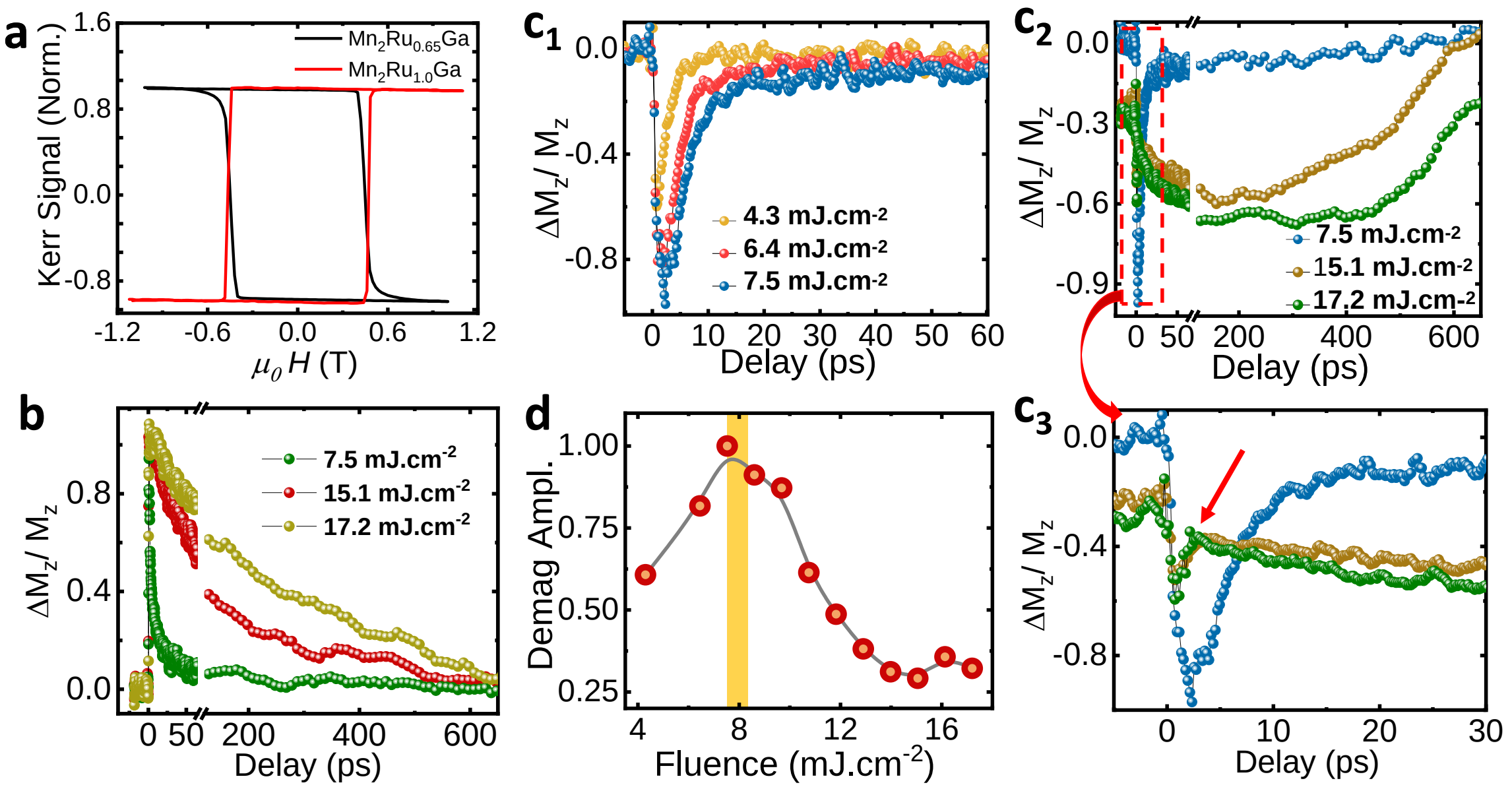

Supplementary Information

# Single pulse all-optical toggle switching of magnetization without Gd: The example of $Mn_2Ru_xGa$

C. Banerjee, N. Teichert, K. Siewierska, Z. Gercsi, G. Atcheson, P. Stamenov,

K. Rode, J. M. D. Coey and J. Besbas*

[1] CRANN, AMBER and School of Physics, Trinity College, Dublin 2, Ireland

*besbasj@tcd.ie

## I. STRUCTURAL AND MAGNETIC CHARACTERIZATION OF $Mn_2Ru_{1.0}Ga$

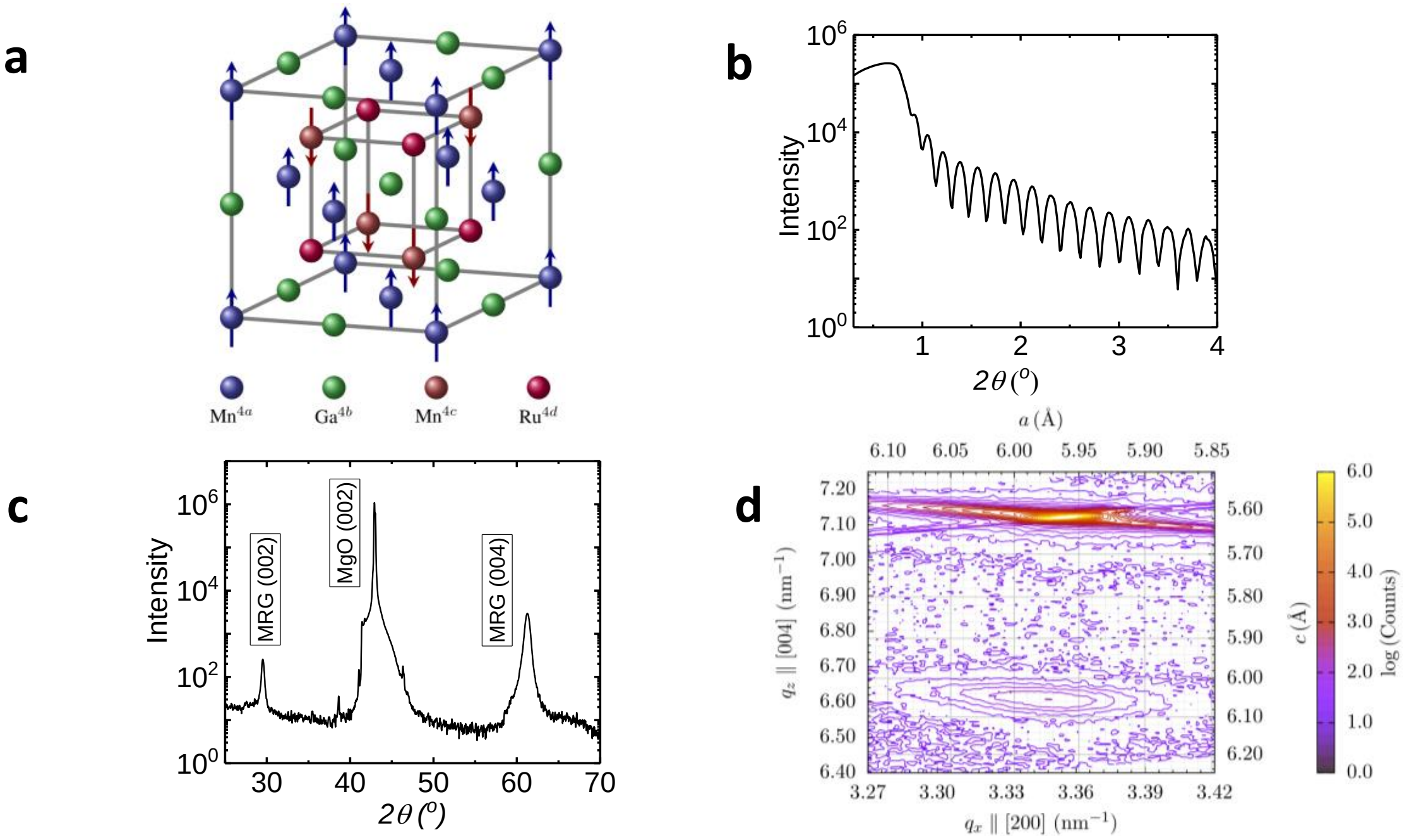

Figure S1: (a) Diagram of the inverted Heusler (XA) crystal unit cell of typical MRG film. Characterisation of $Mn_2Ru_{1.0}Ga$ (MRG): (b) X ray reflectivity pattern of the MRG thin film. Fitting gives a thickness of 42.8 nm and a density of 8.2 g. cm$^{-3}$, (c) X ray diffraction pattern of MRG thin film on MgO (001) substrate. (d) Reciprocal space map of MgO (113) peak and MRG (204) peak with lattice parameters calculated with respect to the MRG unit cell.

MRG crystallises in an inverted Heusler (XA) structure of space group F-43m with two crystallographically inequivalent magnetic Mn atoms at Wyckoff positions 4*a* and 4*c* and Ga and Ru atoms occupy the 4*b* and 4*d* positions, respectively, as shown in Fig. S1 (a). Note the Mn(4c) sublattice is non-centrosymmetric. Due to the biaxial non-volume conserving strain of the MgO substrate the unit cell of MRG is tetragonally distorted and hence the space group is reduced to I-42m. X-ray data on the $Mn_2Ru_{1.0}Ga$ film are shown in Fig. S1 (b), (c) and (d). The X-ray reflectivity (XRR) pattern shown in Fig. S1 (b) has been fitted using X'Pert Reflectivity software and the film thickness was calculated to be 42.8 nm. The X-ray diffraction (XRD) pattern in Fig. S1 (c) exhibits (002) and (004) reflections from the MRG, together with peaks from the MgO substrate. The c-parameter calculated from the (004) reflection is 604.7 pm. A reciprocal space map (RSM) of the MRG film in Fig. S1 (d) confirms the c-parameter obtained from XRD, and shows a distribution of a-parameters around the central value of 595.8 pm, which corresponds to that of MgO. This demonstrates how substrate strain induces a ~ 1% tetragonal elongation of the MRG unit cell since the c/a ratio is 1.01, giving rise to the perpendicular magnetic anisotropy found in all MRG films.

## II. DEPENDENCE OF MAGNETIZATION ON APPLIED MAGNETIC FIELD FOR $Mn_2Ru_{0.65}Ga$ AND $Mn_2Ru_{1.0}Ga$

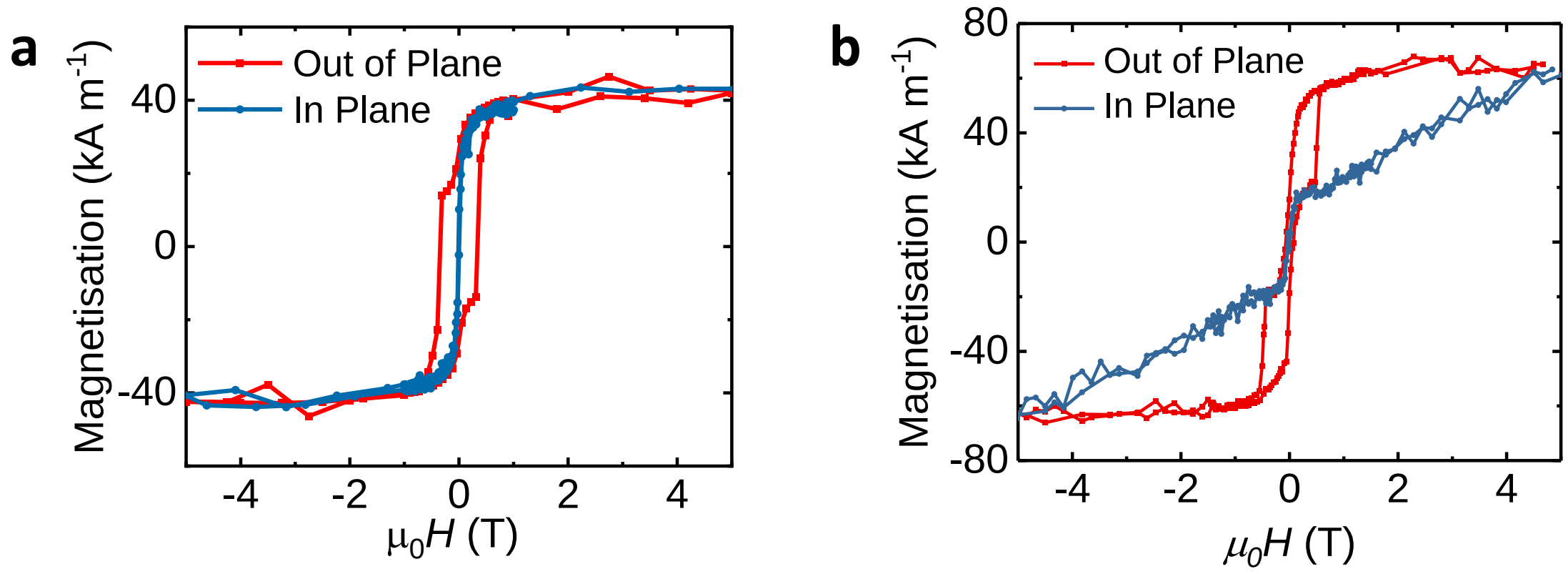

Figure S2: Magnetization versus field applied in plane and out of plane at 300 K (a) for the $Mn_2Ru_{0.65}Ga$ film and (b) for the $Mn_2Ru_{1.0}Ga$ film.

The dependence of magnetization on the field applied out of plane and in plane at 300 K relative to the surface of the thin film was measured for $Mn_2Ru_{0.65}Ga$ and $Mn_2Ru_{1.0}Ga$ films (See Fig. S2), which have $T_{comp}$ below and above RT respectively. From the curves, the saturation magnetizations are 40 and 65 kA . $m^{-1}$, respectively. Interestingly, we observe a soft component in the out-of-plane, as well as the in-plane magnetization, which originates partly from the non-collinearity of the two exchange-coupled antiferromagnetically aligned Mn magnetic sublattices.

III. MEASUREMENT OF OPTICAL AND TRANSPORT MAGNETOMETRY OF $Mn_2Ru_{1.0}Ga$

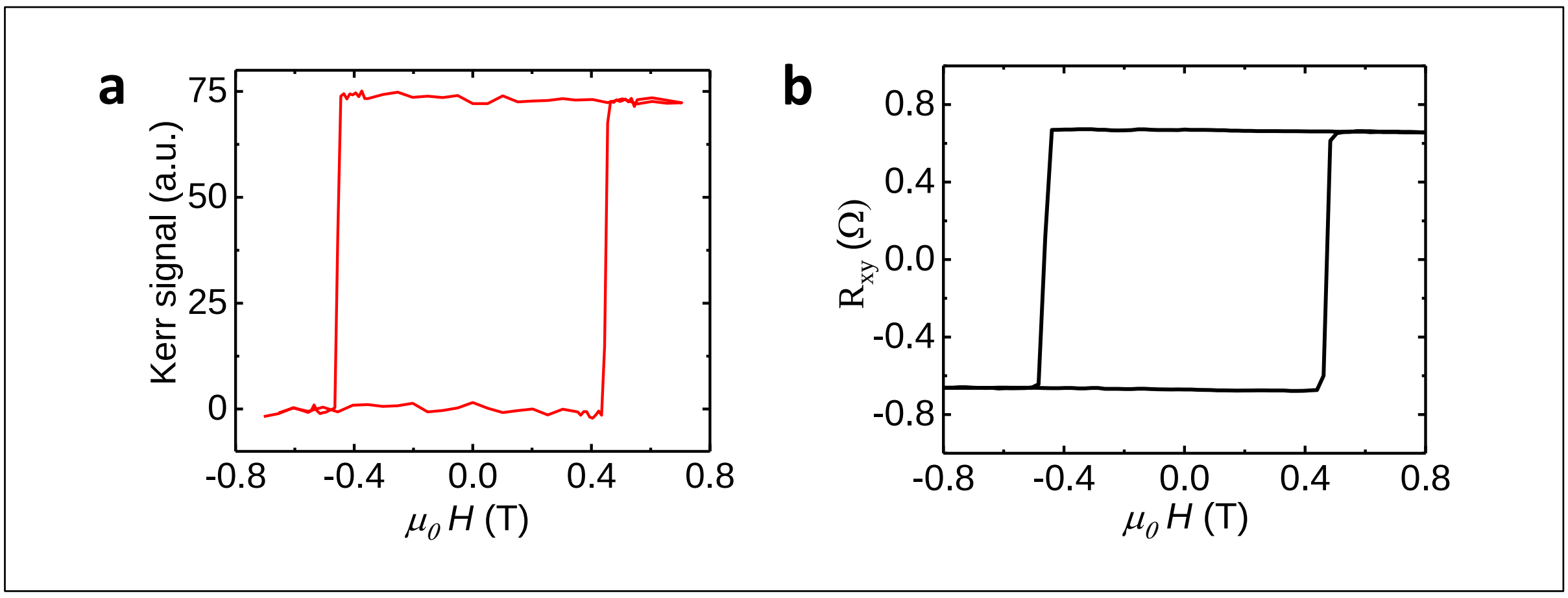

Figure S3: Measured hysteresis loops of $Mn_2Ru_{1.0}Ga$ via (a) Kerr microscopy and (b) anomalous Hall effect.

We have measured the magnetic hysteresis loops using polar Kerr effect (λ = 630 nm) and anomalous Hall effect in the perpendicular $Mn_2Ru_{1.0}Ga$ film, which are compared in Fig. S3. In contrast to the two-step switching observed in the SQUID loop (See Fig. S2 (b)), here the square hysteresis loops exhibit straightforward single-step switching, with an average switching field of 460 mT in both cases. Whereas the SQUID probes the net magnetic moment, the difference of the nearly-equal sublattice contributions, the optical and electrical transport, on the other hand, relies on the distribution of spin-polarized electrons at or near the Fermi energy, which in MRG reflects the 4c sublattice magnetization. Consequently, the hysteresis shown in Fig. S3 as well as the domain images presented here reflects the local magnetization state of the 4c sublattice. It is therefore possible to explore the local magnetization state even at compensation.

## IV. DETERMINATION OF $T_{comp}$ FOR $Mn_2Ru_{1.0}Ga$

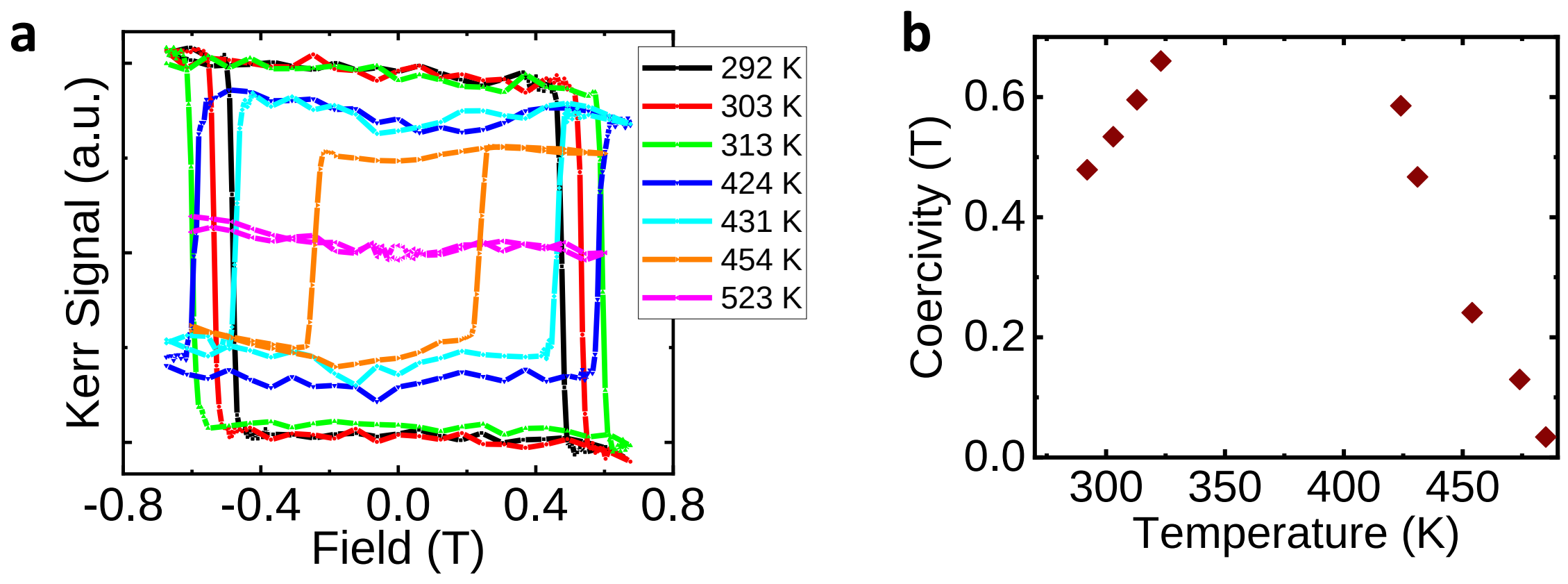

Figure S4: Optically measured hysteresis loops for $Mn_2Ru_{1.0}Ga$ at different temperatures (a) and the corresponding variation of coercive field (b).

Figure S4 (a) presents the hysteresis loops measured by polar Kerr effect in $Mn_2Ru_{1.0}Ga$ at different temperatures. On approaching $T_{comp}$, the net moment falls, which increases the anisotropy field and coercivity, until they diverge at $T_{comp}$. In addition, as the optical measurements probes the Mn(4c) sublattice only, a change in the sign of the hysteresis loop is seen upon crossing $T_{comp}$. The variation of the coercive field as a function of temperature is shown in Fig. S4b, from which $T_{comp}$ for this sample is estimated to be ~ 390 K.

V. RESULT OF PULSE ENERGY DEPENDENT MEASUREMENT IN MRG SAMPLE WHERE AOS WAS NOT OBSERVED

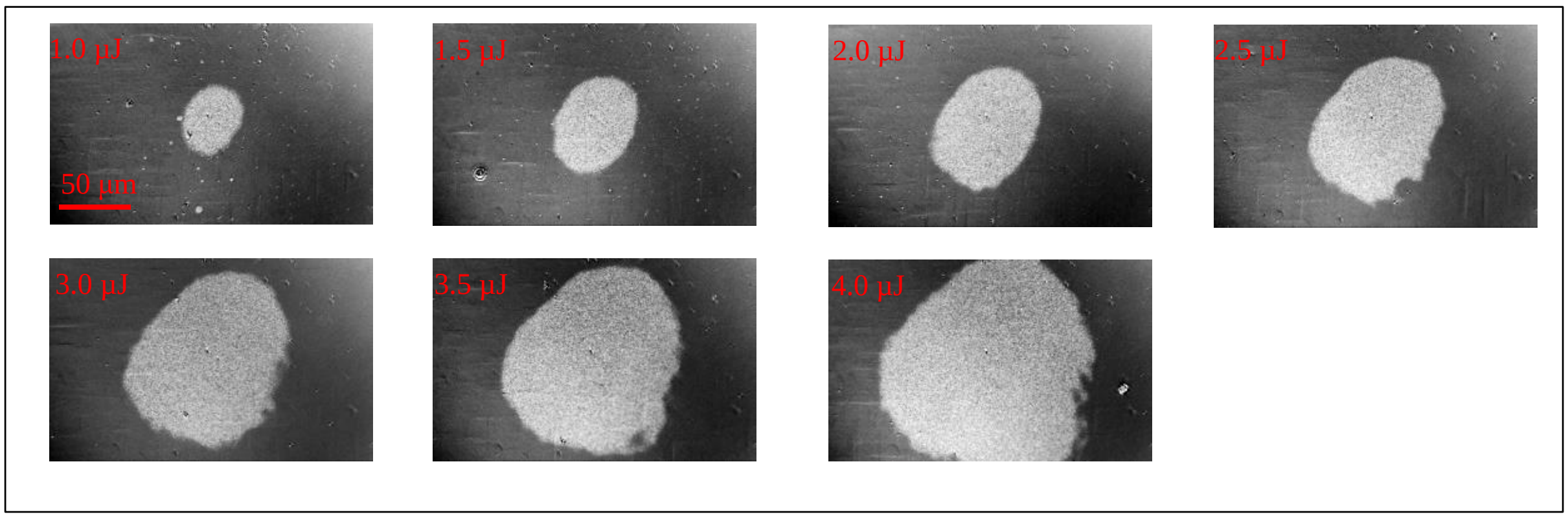

Figure S5: Typical Kerr microscope images of MRG after single laser pulse of various pulse energies were irradiated on different positions of the surface. The results show a multidomain pattern has formed.

In the main text, single-pulse all-optical switching (SP-AOS) is presented for MRG samples having $T_{comp}$ above room temperature. No toggling was not observed for the MRG samples having $T_{comp}$ below room temperature. Figure S5 shows typical Kerr micrographs after the pulse, for different pulse energies. The images reveal the presence of multidomain state with an onset at ~ 1 µJ. This is different to the $Mn_2Ru_{1.0}Ga$ results, where a ring of switched domain was observed around the thermally demagnetized region.

## VI. DETERMINATION OF THE SPOT SIZE AND THRSHOLD FLUENCE FOR SWITCHING

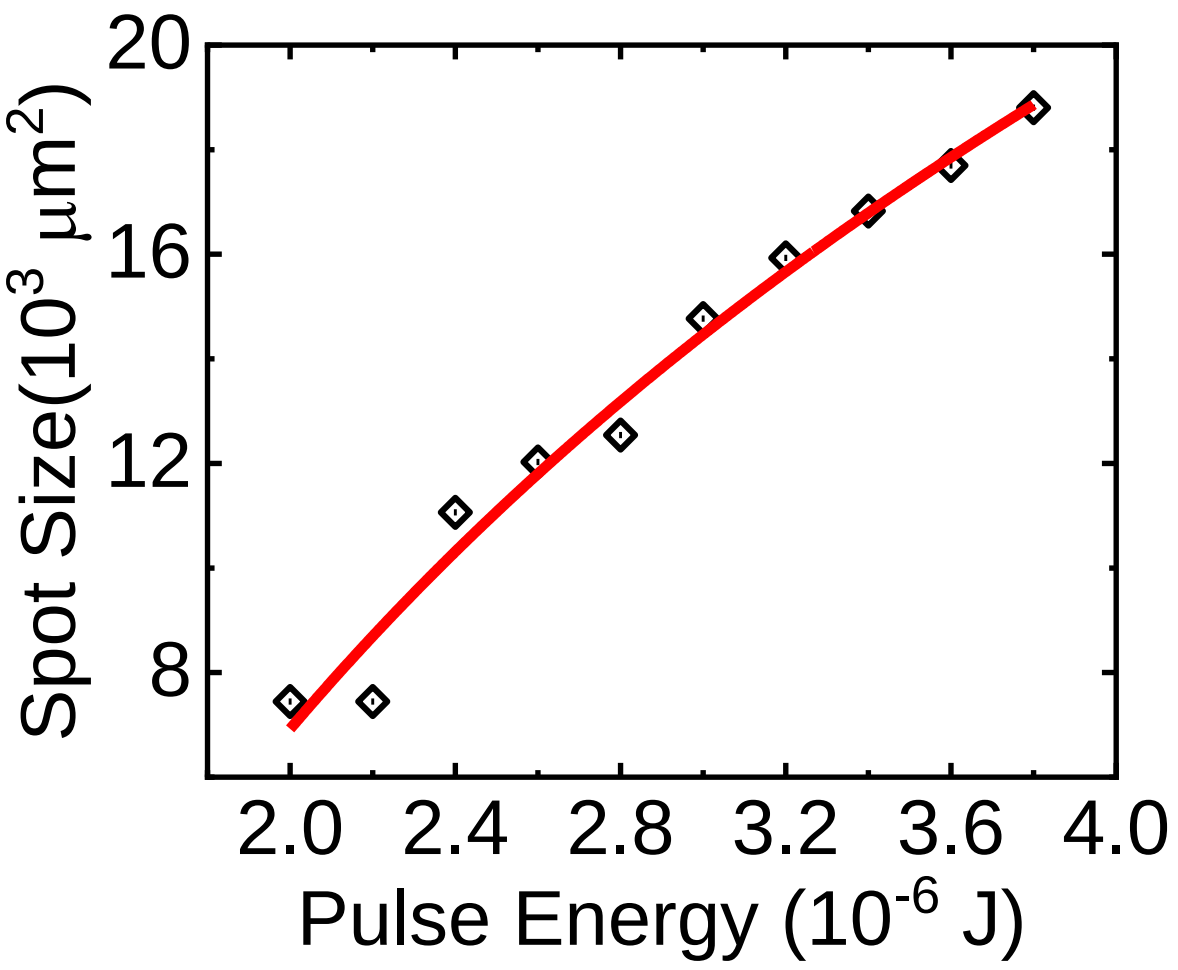

Figure S6: Switched domain size as a function of pulse energy for $Mn_2Ru_{1.0}Ga$. The red solid line is a fit to Eqn. 1.

As mentioned in the main text, we have employed the growth of the switched domain size with increasing pulse energy for $Mn_2Ru_{1.0}Ga$ to calculate the laser spot size as well as the threshold fluence for switching using the Liu method[S1]. This method exploits the fact that at the edge of the magnetic contrast, the excitation fluence is equal to the threshold fluence. By assuming a Gaussian pulse-shape, the switched area can be determined as,

$$A_S = A_0 ln\left(\frac{E}{E_{th}}\right) \qquad \text{.........(1)}$$

where $A_S$ is the switched area corresponding to pulse energy $E$, $A_0$ is the laser spot area and $E_{th}$ is the threshold pulse energy. Figure S6 presents the variation of the switched area as a function of pulse energy, from which the spot area and the threshold pulse energy are extracted by fitting with Eqn. 1 to be 18600 µm$^2$ and 1.4 µJ respectively. The threshold fluence is then calculated by dividing the threshold pulse energy with the laser-spot area ,which yields ~7.5 mJ/cm$^2$.

## VII. SP-AOS MEASUREMENT ON $Mn_2Ru_{0.92}Ga$ USING LASER PULSES OF WAVELENGTH 400 NM

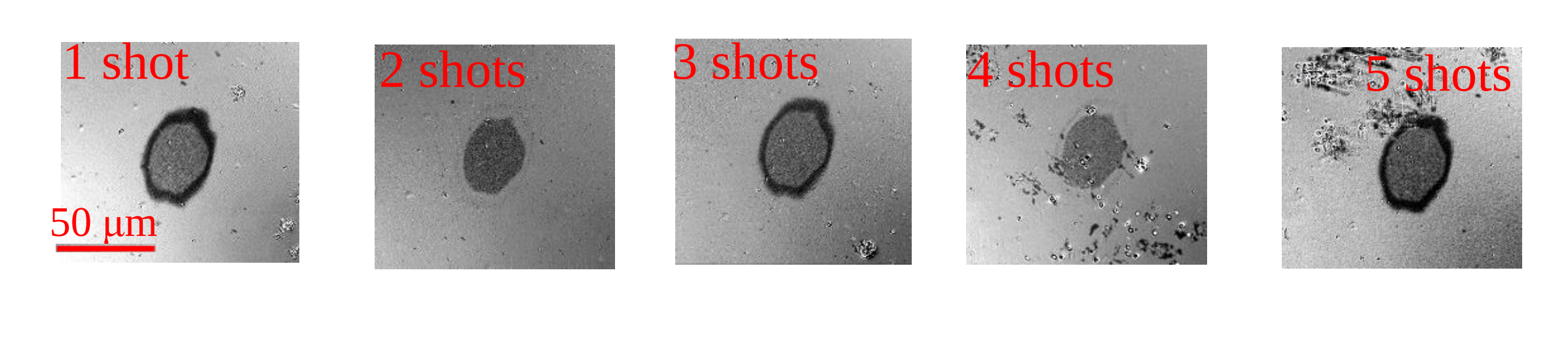

Figure S7: Single pulse all optical switching measurement performed on $Mn_2Ru_{0.92}Ga$ using laser pulses of wavelength 400 nm and pulse energy 0.9 µJ. The number of laser pulses the region is exposed to is labelled in each image.

In order to substantiate the thermal origin of SP-AOS in MRG, we examined the response of its magnetization to the laser pulses of different wavelength and polarization. In Fig. S7 the response of $Mn_2Ru_{0.92}Ga$ is shown for laser light of wavelength 400 nm and up to five laser pulses. The results are in line with the observation with 800 nm (See Fig. 2 in main text). Essentially, a decrease in the excitation wavelength decreases the laser spot size, thereby increasing the thermal gradient across it. This directly affects the magnetization profile of the irradiated region and the switching is observed in a relatively narrow ring around the thermally demagnetized area, as compared to the 800 nm case.

## VIII. COMPARISON OF TEMPERATURE DEPENDENCE OF MAGNETIZATION IN MRG AND GdFeCo

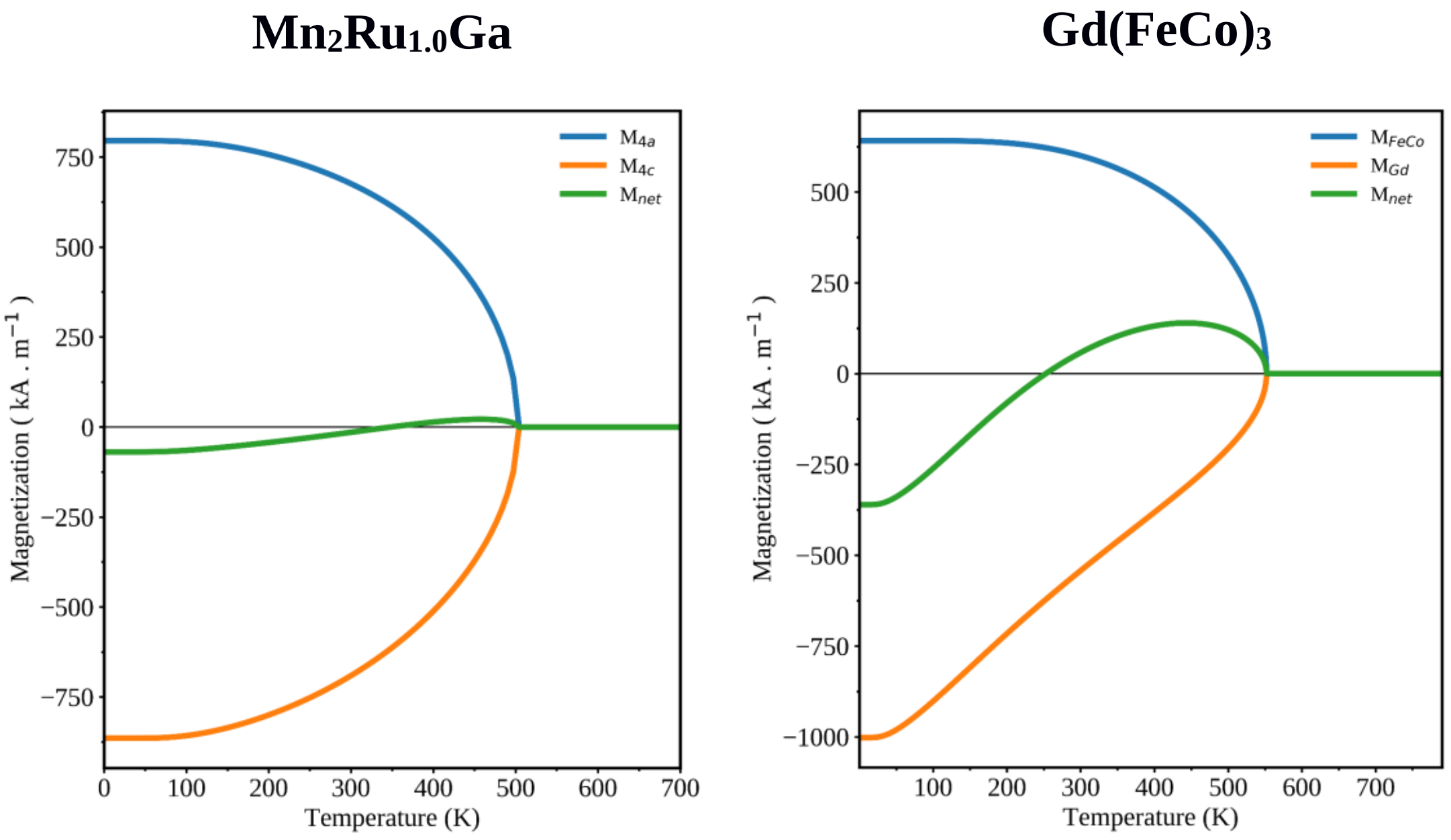

Figure S8: Representation of the sublatttice magnetizations with respect to temperature calculated in the mean field approach for $Mn_2Ru_{1.0}Ga$ and $Gd(FeCo)_3$.

TABLE S1: Outcome of single-pulse excitation in various MRG samples. In some cases, where similar behaviour was found for films with similar composition, only one entry is shown. Values of $T_{comp}$ shown in black are experimentally measured, while the others (shown in blue) are interpolated.

| **Sample** | $T_{comp}$ **(K)** | **Coercive Field (mT)** | **Switching Observed?** |
|---|---|---|---|
| $Mn_2Ru_{0.5}Ga$ | 75 | 150 | No |
| $Mn_2Ru_{0.55}Ga$ | 80 | 170 | No |
| $Mn_2Ru_{0.60}Ga$ | 130 | 260 | No |
| $Mn_2Ru_{0.62}Ga$ | 145 | 350 | No |
| $Mn_2Ru_{0.63}Ga$ | 160 | 370 | No |
| $Mn_2Ru_{0.65}Ga$ | 165 | 440 | No |
| $Mn_2Ru_{0.7}Ga$ | 245 | 740 | No |
| $Mn_2Ru_{0.9}Ga$ | 310 | > 1000 | Yes |
| $Mn_2Ru_{0.92}Ga$ | 315 | > 1000 | Yes |
| $Mn_2Ru_{0.93}Ga$ | 320 | > 1000 | Yes |
| $Mn_2Ru_{0.94}Ga$ | 325 | > 1000 | Yes |
| $Mn_2Ru_{0.95}Ga$ | 375 | 600 | Yes |
| $Mn_2Ru_{1.0}Ga$ | 390 | 480 | Yes |

TABLE S2: MAGNETIC PROPERTIES OF $Mn_2Ru_xGa$ AND $Gd(FeCo)_3$

Magnetic properties of $Mn_2Ru_xGa$ and $Gd(FeCo)_3$ (i.e. GdFeCo). $T_c$ and $T_{comp}$ are respectively the Curie and compensation temperatures. $M_i$ is the magnetization of the sublattice $i$ and $M_{net}$ is the net magnetization of the system. $\tau_i$ are the characteristic demagnetization times for sublattices $i$ = 4a, 4c in MRG and $i$ = Gd, FeCo in $Gd(FeCo)_3$. $\tau_{e-l}$ is the characteristic time associated to the energy transfer between the hot electronic system and the lattice.

| | $Mn_2Ru_xGa$ | $Gd(FeCo)_3$ |
|---|---|---|
| **Structure** | Cubic Heusler XA | Amorphous |
| **$T_c$ (K)** | 500[1] ($Mn_2RuGa$) | 500[6] ($Gd_{22}Fe_{9.8}Co_{68.2}$) |
| **$T_{comp}$ (K)** | 390[1,2] ($Mn_2Ru_{1.0}Ga$) | 250[5] ($Gd_{25}Fe_{65.6}Co_{9.4}$) |
| **$M_{4a/FeCo}$ (kA . $m^{-1}$)** | 790[1] ($Mn_2Ru_{1.0}Ga$)<br>550[3] ($Mn_2Ru_{0.61}Ga$) | 640[7] ($Gd_{25}Fe_{65}Co_{10}$) |
| **$M_{4c/Gd}$ (kA . $m^{-1}$)** | 860[1] ($Mn_2Ru_{1.0}Ga$)<br>590[3] ($Mn_2Ru_{0.61}Ga$) | 1000[7] ($Gd_{25}Fe_{65}Co_{10}$) |
| **$M_{net}$ (kA . $m^{-1}$)** | 70[2] (0 K) ($Mn_2Ru_{1.0}Ga$)<br>40[3] (0 K)($Mn_2Ru_{0.61}Ga$) | 360[7] (0 K) ($Gd_{25}Fe_{65}Co_{10}$) |
| **$\tau_{4a/Gd}$ (ps)** | 8.0[4] ($Mn_2Ru_{0.7}Ga$) | 0.43[5] ($Gd_{25}Fe_{65.6}Co_{9.4}$) |
| **$\tau_{4c/FeCo}$ (ps)** | 0.5 ($Mn_2Ru_{1.0}Ga$),<br>0.5/8.0[4] ($Mn_2Ru_{0.7}Ga$) | 0.1[5] ($Gd_{25}Fe_{65.6}Co_{9.4}$)<br>4.0/150[6] ($Gd_{22}Fe_{9.8}Co_{68.2}$) |
| **$\tau_{e-l}$ (ps)** | 2.0[4] ($Mn_2Ru_{0.7}Ga$) | 1.5[5] ($Gd_{25}Fe_{65.6}Co_{9.4}$)<br>5.0[6] ($Gd_{22}Fe_{9.8}Co_{68.2}$) |

1 From static MOKE vs Temperature measurements.
2 From SQUID/XRD measurements.
3 Fowley *et al. Phys. Rev. B* **98**, 220406(R) (2018).
4 Bonfiglio *et al.* ArXiv, 2003.01420 (2020).
5 Radu *et al.* Nature **472**, 205-209 (2011).
6 Mekonnen *et al.* Phys. Rev. B **87**, 180406(R) (2013).
7 Estimated for atomic spins $\mu_{Gd}$ =7.6 $\mu_B$ and $\mu_{Fe/Co}$ =1.6 $\mu_B$.